# Accelerating Discovery of Vacancy Ordered 18-Valence Electron Half-Heusler Compounds: A Synergistic Approach of Machine Learning and Density Functional Theory

Gowri Sankar S[1,2], Mukesh K. Choudhary[1,2], Amal Raj V[2] and P. Ravindran, [a)1,2]

[1]*Department of Physics, Central University of Tami Nadu, Thiruvarur, Tamil Nadu, India.*
[2]*Simulation Centre for Atomic and Nanoscale MATerials (SCANMAT), Central University of Tamil Nadu, Thiruvarur, Tamil Nadu 610101, India*

[a)] Corresponding author: raviphy@cutn.ac.in

**Abstract.** In this study we attempted to model vacancy ordered half Heusler compounds with 18 valence electron count (VHH) derived from 19 VEC compounds such as TiNiSb such that the compositions will be $Ti_{0.75}NiSb$, $Zr_{0.75}NiSb$ and $Hf_{0.75}NiSb$ with semiconducting behavior. The main motivation is that such a vacancy ordered phase not only introduce semi conductivity but also it will disrupt the phonon conducting path in HH alloys and thus reduce the thermal conductivity and as a consequence enhance the thermoelectric figure of merit. In order to predict the formation energy ($\Delta H_f$) from composition and crystal structure we have used 4684 compounds for their $\Delta H_f$ values are available in the material project database and trained a machine learning model with $R^2$ value of 0.943. Using this trained model, we have predicted the $\Delta H_f$ of a list of VHH. From the predicted database of VHH we have selected $Zr_{0.75}NiSb$ and $Hf_{0.75}NiSb$ to validate the machine learning prediction using accurate DFT calculation. The calculated $\Delta H_f$ for these two compounds from DFT calculation are found to be comparable with our ML prediction. The calculated electronic and lattice dynamics properties show that these materials are narrow band gap semiconductors and are dynamically stable as their all-phonon dispersion curves are having positive frequencies. The calculated Seebeck coefficient, electrical conductivity as well as thermal conductivity, power factor and thermoelectric figure of merit are analyzed.

## INTRODUCTION

Machine Learning (ML) serves various purposes in numerous fields, such as business, medical sciences, and agriculture, among others. The continuous development of advanced algorithms and the availability of vast amount of data have significantly enhanced the accuracy and robustness of ML models. However, further progress is still ongoing to create field-specific or target/property-specific models those can achieve desirable accuracy [1]. Despite the abundance of data, a considerable portion remain unstructured and unclean, making it time-consuming to filter out relevant information. Acquiring high-quality data from this unfiltered dataset is crucial for building successful models. Additionally, the quantity and quality of data used for training the model plays a significant role. In essence, an ideal dataset should be specific to the targeted property, have sufficient quantity, and encompass various possible variations to have unbiased prediction. Unfortunately, obtaining an ideal dataset often becomes a bottleneck, which consequently affects the predictive capacity of the model, regardless of the effectiveness of the ML algorithm. Inconsistencies in the data can also undermine the model's performance. In the context of materials science, researchers are continuously exploring new dataset repositories and employing new methods of feature importance analysis. Predicting a potential material with targeted property from a wide chemical space holds particular importance in green energy technologies and other applications. To achieve accurate predictions, several factors come into play, including the choice of ML algorithm, dataset quality, feature selection, and model validation. The dataset should be representative, clean, error-free, and contain relevant features. Understanding basic physical, chemical, and thermodynamical properties can help narrow down the search for essential features and guide the selection of ML models. However, the relationship between the target property and descriptors might not be fully known and could be complex to model. In such cases, trying out various models to achieve the required accuracy may be the only solution. Additionally, another challenge could be the lack of effective descriptors. Addressing these challenges and considering these factors, researchers can enhance the capabilities of ML in predicting materials with specific properties and other domains effectively. In this study we have considered the HH compounds which are ternary intermetallic compounds those have general formula *XYZ*, where *X* and *Y* are typically transition metals, and *Z* is a main group element. The semiconducting compounds based on HH can convert waste heat energy into useful electrical energy, making them ideal for thermoelectric applications. The transfer of electrons from the electropositive element *X* to more electronegative elements *Y* and *Z* provides a stable closed-shell configuration if they fulfil 18 VEC. This results in semiconducting behaviour, which is essential for a high efficiency thermoelectric material. In the case of converting a 19 VEC system into an 18 VEC system we can use point defects such as vacancies or substitutional to remove one of the electrons. The present study we have introduced ordered vacancies in the *X* site to bring 18 VEC to the system so that the semiconducting behaviour will appear and thermal conductivity will reduce such that the resulting material will have high thermoelectric figure of merit (*ZT*).

## HYBRID ML+DFT APPROACH

In this study, we have discussed how to achieve accurate predictions for the $\Delta H_f$ of HH with ordered vacancies those possess 18 VEC, specifically focusing on the importance of feature engineering in ML methods. Several studies have been done for calculating the $\Delta H_f$ through ML approach. Mao *et al.* [2] has studied $\Delta H_f$ of binary compounds using ML techniques and achieved $R^2$ value of 0.94 on test data. Rengaraj *et al.* [3] used ML approach for predicting $\Delta H_f$ for ternary compounds using a two-step ML approach, in first step used classifier and in second step they used regression method and achieved the MAE value of 0.129 eV/atom. Selecting and engineering relevant features for a given dataset is critical for achieving the best model accuracy. To simplify the process, we used the featurization method provided in "Matminer",[4] which incorporates both composition and crystal structure information of the materials for featurisation. We adapted 10 regressor ML models, namely CatBoost, Random Forest, Bagging, AdaBoost, Decision Tree, Light Gradient Boosting Machine, Gradient Boost, eXtreme Gradient Boost, Linear, and Lasso for the present study. The input features were selected through analysis of heat map plotted with correlation among features and also targeted property vs features were selected using Spearman, Kendall, and Pearson correlation methods [5]. Through analysis of the correlation heatmaps from these three methods, we identified common features in them which show low correlation between the features-target and importance of feature for targeted property. Correlation values are having the range from -1 to 1, where a high absolute value indicating a stronger correlation between features. In case of features correlation matrix, features with low or no correlation with other features were retained, as they were likely to provide essential information to the model and in feature-target correlation matrix with zero correlation values were dropped, finally giving us 35 features for our model training.

We also considered feature importance with permutation to finalize the essential feature set. By employing these methods, we effectively reduced the dimensionality of our feature space while retaining the most informative features for training our predictive model. These feature selection and processing methods played a critical role in deciding the accuracy of our trained model. To assess our model performance, we utilized two standard metrics: RMSE (Root Mean Square Error) and $R^2$ (coefficient of determination). Lower RMSE values indicate better model performance, while $R^2$ measures how well the data fits the regression line, with values typically ranging from 0 to 1. Our analysis and discussion provide valuable insights into accurately predicting $\Delta H_f$ of VHH using ML techniques.

We used Material Project [6] (MP) API to acquire data from MP database for 4684 ABC- type compositions, their structure, and corresponding $\Delta H_f$ values obtained from GGA calculations. The structure and composition features obtained from Matminer library was used to create features to train all our models. We split the dataset into training

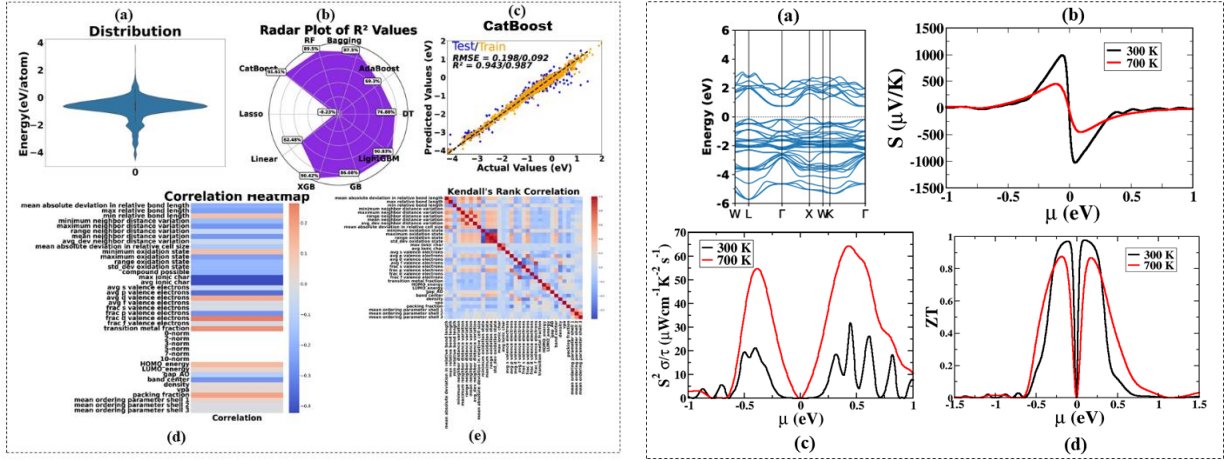

**Figure. 1** (left panel). (a) Distribution of $\Delta H_f$ per atom values in the dataset.,(b) The $R^2$ value from 10-fold cross validation with 10 ML models considered in the present study, (c) Performance of CatBoost model, (d) Heatmap of target-features correlation and (e) Heatmap of correlation among features. **Figure. 2** (right pannel) The calculated electronic structure and transport properties of $Zr_{0.75}NiSb$. a) electronic structure (b) Seebeck coefficient (c) power factor and (d) Figure of merit as a function of chemical potential $\mu$ in the range of −1.5 to 1.5 eV at 300 and 700 K obtained from PBE-GGA calculation.

and testing set with 8:2 ratio. Tenfold cross validation was done for the above mentioned 10 regression models based on $R^2$ value as shown in Fig. (1) (b). Among the 10 ML models considered in the present study CatBoost regressor gave a $R^2$ value of 0.916 closely followed by LGBM and XGB with the $R^2$ value of 0.908 and 0.904 in cross validation, respectively. From the cross validation we found that CatBoost is the best model among them and hence it was further trained on the entire data to achieve increased performance. Using the trained model, we have predicted formation energy of vacancy ordered HH with valence electron count 18 having composition $X_{0.75}YZ$ and from that identified several compounds having high stability. It may be noted that Anand *et al.*[7] used the convex hull method to predict the stable phases of VHH and found the compositions such as $Ti_{0.75}NiSb$, $Zr_{0.75}NiSb$, and $Hf_{0.75}NiSb$ for that some experimental verifications are also made. Our trained model also correctly predicted these materials as stable compounds with negative formation energy as given in Table.1

The DFT calculations for structural optimization and electronic structure were performed using the Vienna ab-initio simulation package [8] (VASP) with the PAW method [9]. The PBE-GGA [10] was used for the exchange-correlation potential to compute equilibrium structural parameters. Monkhorst pack [11] scheme was used to sample the irreducible part of the first Brillouin zone (IBZ) with a $12 \times 12 \times 12$ k mesh for geometry optimization. Geometry optimization utilized a plane-wave energy cutoff of 600 eV with an energy convergence criterion of $10^{-6}$ eV/cell and a force convergence criterion of less than 1 meV/Å. Previous studies have shown that these computational parameters accurately predict the equilibrium structural parameters for half Heusler alloys [12]. For electronic structure calculations, high symmetry directions of the first IBZ of FCC lattice was considered.

**Table 1**. The formation energy ($\Delta H_f$) predicted by ML and DFT for selected compounds and their corresponding PBE-GGA band gap values.

| Compounds | ML predicted (eV/atom) | DFT calculated (eV/atom) | PBE-GGA (eV) |
|---|---|---|---|
| $Ti_{0.75}NiSb$ | -0.42 | -0.86 | 0.66 |
| $Hf_{0.75}NiSb$ | -0.34 | -0.684 | 0.77 |
| $Zr_{0.75}NiSb$ | -0.62 | -0.689 | 0.74 |
| TiCoSb | -0.270 | -0.659 | 1.04 |
| KOH | -1.45 | -1.545 | 3.68 |
| NaOH | -1.62 | -1.449 | 2.80 |
| ZrNiSb | -0.384 | -0.459 | Metal |
| TiNiSb | 0.18 | -0.79 | Metal |
| VCoSb | -0.288 | -0.189 | Metal |

All the 3 VHH compounds considered in the present study exhibit an indirect band gap behavior and the representative system $Zr_{0.75}NiSb$ band structure is shown Fig. 2 (a) with the valence band maximum located at the *X* point, and the conduction band minimum located at the *Γ* point and the corresponding band gap values obtained from PBE-GGA calculations are given in Table 1. Our calculated bandgap value for $Ti_{0.75}NiSb$ is 0.66 eV, which agrees well with previous report [13]. The calculated band gap values for $Zr_{0.75}NiSb$ and $Hf_{0.75}NiSb$ are 0.75 eV and 0.77 eV, respectively. We used the Phonopy code [14] with the PBE-GGA functional to calculate the phonon dispersion curves for these three VHH compounds to check their dynamic stability. The finite difference method calculations were done at equilibrium volume using a $3 \times 3 \times 3$ supercell. The calculated phonon band structure show that these compounds are dynamically stable at ambient conditions, as there were no negative frequencies in their phonon dispersion curves. We have also calculated the TE transport properties such as Seebeck coefficient (Fig.2 (b)), power factor (Fig. 2(c)), and *ZT* (Fig. 2 (d)). Fig. 2(b) shows the Seebeck coefficients of $Zr_{0.75}NiSb$ as a function of chemical potential ranging from -1 eV to 1 eV, at 300/ 700 K. Interestingly, the Seebeck coefficient values were nearly twice as high at room temperature compared to those at 700K for $Zr_{0.75}NiSb$. The calculated Seebeck coefficient for $Zr_{0.75}NiSb$ at 300 K is 1000 µV/K, and at 700 K it reduces to 456 µV/K. The calculated electrical conductivity decreased from $2.96 \times 10^{20}$ (1/Ωms) at 300 K to $2.73 \times 10^{20}$ (1/Ωms) at 700 K. Fig. 2 (c) shows the power factor of $Zr_{0.75}NiSb$ and that exhibits a maximum value of 64 µW/ $K^2$cms at 700 K for µ = 0.42 eV and a maximum value of 32 µW/ $K^2$cms at 300 K for µ = 0.44 eV, indicating efficient heat-to-electricity conversion will happen at elevated temperatures. The corresponding *ZT* value is 0.97 and 0.87 at 300 K and 700 K, respectively suggesting that this material has potential to use in high temperature thermoelectric applications.

## CONCLUSION

In conclusion, our aim was to predict the formation energy of VHH compounds using ML model and validate the prediction using accurate DFT calculations. To accomplish this, we utilized a dataset comprising 4684 ABC-type compounds, which we obtained from the Material Project. To determine the best performing model among the ten ML models, we employed cross-validation. Among them, CatBoost emerged as the top performer, exhibiting an impressive $R^2$ value of 0.916. For further evaluation, we split the data into a training set and a test set in an 8:2 ratio, and the CatBoost model achieved excellent results with a test $R^2$ value of 0.943 and an RMSE value of 0.198. The cross-validation approach proved to be invaluable in selecting the most accurate model for our specific task. Additionally, we took great care in enhancing our ML model's performance by judiciously selecting appropriate features for training. To do this, we analyzed correlation matrices to identify the most relevant features. Moreover, we carried out feature dimensionality reduction to optimize the model's performance. By combining DFT and ML approaches, we demonstrated how this synergistic approach accelerates the discovery of potential new materials by searching wide chemical space, thus contributing to the advancement of materials science research. We also found that creating VHH by introducing vacancy defects in 19 VEC HH compounds one can create semiconducting compounds with tunable bandgap values. Additionally, we have reported the electronic structure, Seebeck coefficient, power factor and figure of merit for VHH $Zr_{0.75}NiSb$. This defect engineering approach is expected to plays an important role in the development of higher efficient thermoelectric materials for waste heat recovery and thermoelectric refrigeration.

## ACKNOWLEDGMENTS

The authors are grateful to the SCANMAT Centre, Central University of Tamil Nadu, Thiruvarur for providing the computer time at the SCANMAT supercomputing facility**.** The authors are also grateful to the Science and Engineering

Research Board (SERB), India, for funding support under the project SERB-Core Research Grant (CRG) vide file no. CRG/2020/001399.## REFERENCES

Research Board (SERB), India, for funding support under the project SERB-Core Research Grant (CRG) vide file no. CRG/2020/001399.

## REFERENCES

1. Chen, Chi, et al. "A critical review of machine learning of energy materials." *Advanced Energy Materials* 10.8 (2020): 1903242.
2. Mao, Yuanqing, *et al*. "Prediction and classification of formation energies of binary compounds by machine learning: an approach without crystal structure information." *ACS omega* 6.22 (2021): 14533-14541.
3. Rengaraj, Varadarajan, *et al.* "A Two-Step Machine Learning Method for Predicting the ΔHf of Ternary Compounds." Computation 11.5 (2023): 95.
4. Ward, Logan, *et al.* "Matminer: An open-source toolkit for materials data mining." Computational Materials Science 152 (2018): 60-69.
5. van den Heuvel, Edwin, and Zhuozhao Zhan. "Myths about linear and monotonic associations: Pearson's r, Spearman's ρ, and Kendall's τ." The American Statistician 76.1 (2022): 44-52.
6. A. Jain, S. P. Ong, *et al*., Commentary:The materials project: A materials genome approach to accelerating materials innovation, APL materials 1 (1) (2013) 011002
7. Anand, Shashwat, *et al*. "A valence balanced rule for discovery of 18-electron half-Heuslers with defects." Energy & Environmental Science 11.6 (2018): 1480-1488.
8. G. Kresse, J. Furthmüller, Efficiency of ab-initio total energy calculations for metals and semiconductors using a plane-wave basis set,Computational materials science 6 (1) (1996) 15–50.
9. G. Kresse, D. Joubert, from ultrasoft pseudopotentials to the projector augmented-wave method, Physical review b 59 (3) (1999) 1758
10. J. P. Perdew, K. Burke, M. Ernzerhof, generalized gradient approximation made simple, Physical review letters 77 (18) (1996) 3865.
11. H. J. Monkhorst, J. D. Pack, Special points for brillouin-zone integrations, Physical review B 13 (12) (1976) 5188
12. Choudhary, M. K., & Ravindran, P. (2022). First principle design of new thermoelectrics from TiNiSn based pentanary alloys based on 18 valence electron rule. Computational Materials Science, 209, 111396.
13. Luo, Feng, et al. "18-Electron half-Heusler compound Ti 0.75 NiSb with intrinsic Ti vacancies as a promising thermoelectric material." Journal of Materials Chemistry A 10.17 (2022): 9655-9669.
14. A. Togo, I. Tanaka, First principles phonon calculations in materials science, Scripta Materialia 108 (2015) 1–5